\documentclass[a4paper,11pt]{article}
\usepackage{pos}

\title{A study of systematic uncertainties within the MSHT PDF Framework}

\author*[a]{Matthew Reader}

\affiliation[a]{University College London,\\
  Gower Street, London, WC1E 6BT, United Kingdom}

\emailAdd{matthew.reader.21@ucl.ac.uk}

\abstract{Experimental errors are now incredibly precise, and are often dominated by the systematic uncertainties.  Therefore the errors obtained in the Parton Distribution Functions that are extracted from this data will also be dominated by these experimental systematic errors, as well as  the systematic errors embedded in the theoretical calculations. However, as is well known, there are often significant uncertainties in these systematic errors, and so to determine precisely the errors in the Parton Distribution Functions, we need to be thoughtful about the uncertainties in the errors themselves.  In this paper, we discuss an approach where these "errors on errors" can be incorporated into a $\chi^2$ calculation, and investigate how such a model behaves and what it tells us about the resulting errors. Also we look at two data sets, ATLAS W,Z Data \cite{ATLAS:2016nqi} and the ATLAS 7 TeV Inclusive Jet Distribution Data \cite{ATLAS:2014riz} and investigate the information that this model implies about these two data sets.}

\FullConference{31st International Workshop on Deep Inelastic Scattering (DIS2024)\\
 8–12 April 2024\\
Grenoble, France\\}

\begin{document}
\maketitle
\vspace{-0.3cm}
\section{Introduction}
\vspace{-0.3cm}
Experimental errors are becoming extremely precise and are now dominated by systematic uncertainties. However, there are often significant errors in the determination of these systematic errors. Therefore, it is becoming increasingly important that these "errors on errors" are incorporated into our Parton Distribution framework such that the extracted errors incorporate this extra layer of uncertainty. In this short document, we demonstrate how it is possible to achieve this.

\vspace{-0.3cm}
\section{Derivation of the Model}
\vspace{-0.3cm}
Consider a set of data, $\textbf{y}$. The probability of $\textbf{y}$ can be written $P(\textbf{y} | \mu,\theta)$, where $\mu$ are parameters of interest and $\theta$ are nuisance parameters that are required for the correctness of the model.
If we let $\theta = (\theta_1,...,\theta_N)$ be independent Gaussian distributed values $u = (u_1,...,u_N)$, with  standard deviations $\sigma_u = (\sigma_{u_1}...,\sigma_{u_N})$, then the Likelihood function can be written as:
\begin{equation*}
   L(\mu,\theta)= P(\textbf{y},\textbf{u}|\mu,\theta) = P(\textbf{y}|\mu,\theta)P(\textbf{u}|\theta)
\end{equation*}  
\begin{equation}
    = P(\textbf{y} | \mu,\theta) \prod_{i=1}^N \frac{1}{\sqrt{2 \pi} \sigma_{u_i}} e^{-(u_i-\theta_i)^2/2\sigma_{u_i}^2}
\label{Start Eqn}    
\end{equation}
However, $\sigma_{u_i}$ maybe uncertain. One way to incorporate this uncertainty in $\sigma_{u_i}$ has been proposed in \cite{Cowan:2018lhq}. In this proposal we model the estimated variances, $v_i$, of $\sigma^2_{u_i}$, as Gamma distributed, which allows us to rewrite equation \ref{Start Eqn} as:
\begin{equation}
\hspace{-0.5cm}
    L(\mu,\theta,\sigma_{u_i}^2) = P(y|\mu,\theta) \prod^N_{i=1} \frac{1}{\sqrt{2\pi}\sigma_{u_i}}e^{-(u_i-\theta_i)^2/2\sigma_{u_i}^2} \frac{\beta_i^{\alpha_i}}{\Gamma(\alpha_i)}v_i^{\alpha_i-1}e^{-\beta_i v_i}
    \label{gamma model}
\end{equation}
where $\alpha_i = 1/(4r_i^2)$, $\beta_i = 1/(4r_i^2 \sigma_{u_i}^2)$ and $r_i$ is defined as the relative uncertainty in the estimate of the systematic error. The parameters $r_i$ can therefore be referred to as the "error on errors".
This model can be reinterpreted as a Student's t-distribution, once we make a small change of variables:
\begin{equation}
L(\mu,\theta,\sigma_{u_i}^2) = P(y|\mu,\theta) \prod^N_{i=1} \frac{\Gamma(\frac{\nu_i+1}{2})}{\sqrt{\nu_i\pi} \Gamma(\nu_i /2)}\left(1+\frac{t_i^2}{\nu_i}\right)^{-\frac{\nu_i+1}{2}}
\end{equation}
where $t_i = \frac{u_i-\theta_i}{\sqrt{v_i}}$ and $\nu_i=\frac{1}{2r_i^2}$. Therefore,  we can treat our nuisance parameters as t-distributed!

Now we want to extend this model to incorporate correlated systematic errors. So consider:
\begin{equation}
    y_i = d_i + errors = d_i + \sigma_i z_i + \sigma_{u_i} t_{u_i}+ \sum_{j=1}^{M} \beta_{ij}t'_{j}
\end{equation}
where for each observable $y_i$ we have one statistical error $\sigma_i$, with a $z_i$ that is a Normally distributed fluctuating variable, one uncorrelated systematic error $\sigma_{u_i}$  with a $t_{u_i}$ that is a t-distributed fluctuating variable with dof of $\nu=1/2r_{\chi^2}^2$, and M correlated systematic errors, $\beta_{ij}$, each with a fluctuation $t'_j$ that are t-distributed with degree of freedom of  $\nu=1/2r_{\chi^2}^2$. 

If we treat all the t-distributions as independent
\footnote{Note that if we treated the t-distributions as a Multi-variate t-distribution with zero correlation between all the $t_u$ and the $t'$, then the likelihood function would be different.}, then the Log-likelihood function, once we have maximized with respect to $z_i$, can be written up to some constants as:

\begin{equation*}
  -2LnL =  \sum_{i=1}^N \left(\frac{m_i-d_i-\sigma_{u_i}t_{u_i}-\sum_j \beta_{ij}t'_j}{\sigma_i}\right)^2  
\end{equation*}
\begin{equation}
    + (\nu+1)\sum_{i=1}^N Ln  \left(1+\frac{t_{u_i}^2}{\nu} \right)+(\nu+1) \sum_{j=1}^M  Ln   \left(1+\frac{{t'}_{j}^2}{\nu} \right)    \equiv \chi^2
    \label{chi2_1 ru and rj t-dist}
\end{equation}
where we can define this to be a $\chi^2$ once we have minimized simultaneously with respect to both $t_{u_i}$ and ${t'}_j$.

\vspace{-0.3cm}
\section{The case of Statistical and uncorrelated systematic errors only}
\label{Section 2}
\vspace{-0.3cm}
Let's initially consider the case of only statistical and uncorrelated systematic errors. In this case we can write $y_i = d_i +\sigma_i z_i + \sigma_{u_i} t_{u_i}$, where $z_i \sim N(0,1)$, $t_{u_i}\sim t(0,\nu=1/2r^2_{Dist})$. That is, we are drawing our $y$ from a distribution where the statistical errors are normally distributed and the uncorrelated systematic errors are t-distributed.

\begin{figure}[!htb]
    \centering
    \begin{minipage}{.45\linewidth}
        \centering
        \includegraphics[width=1.\linewidth, height=0.25\textheight]{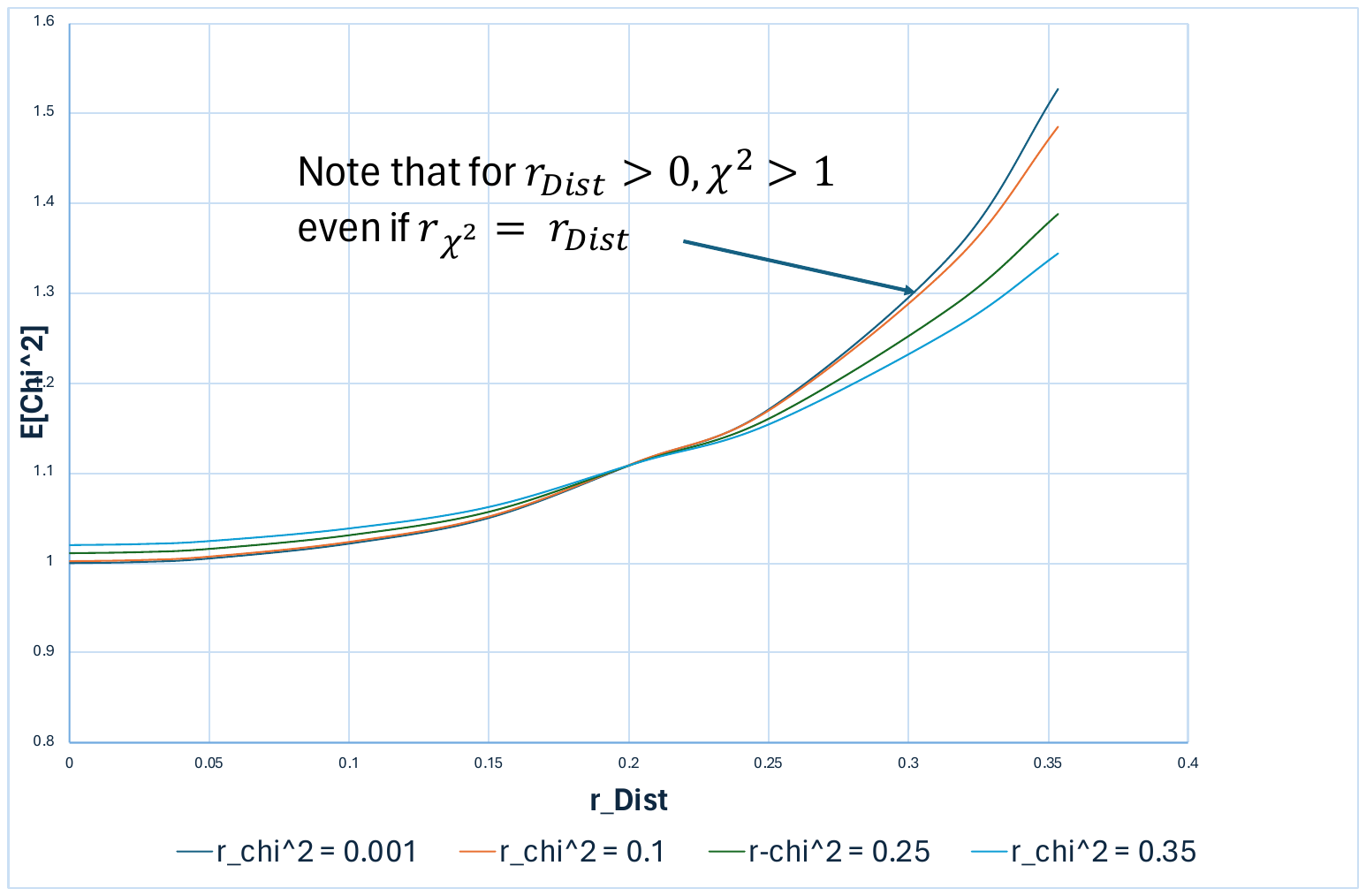}
        \caption{Graph of $E[\chi^2]$ as a Function of $r_{Dist}$ for 4 different $r_{\chi^2}$ ($\sigma_{i}=\sigma_{u_i} = 1 $)}
        \label{Figure 1}
    \end{minipage}\hfill
    \begin{minipage}{0.45\textwidth}
        \centering
        \includegraphics[width=1.\linewidth, height=0.25\textheight]{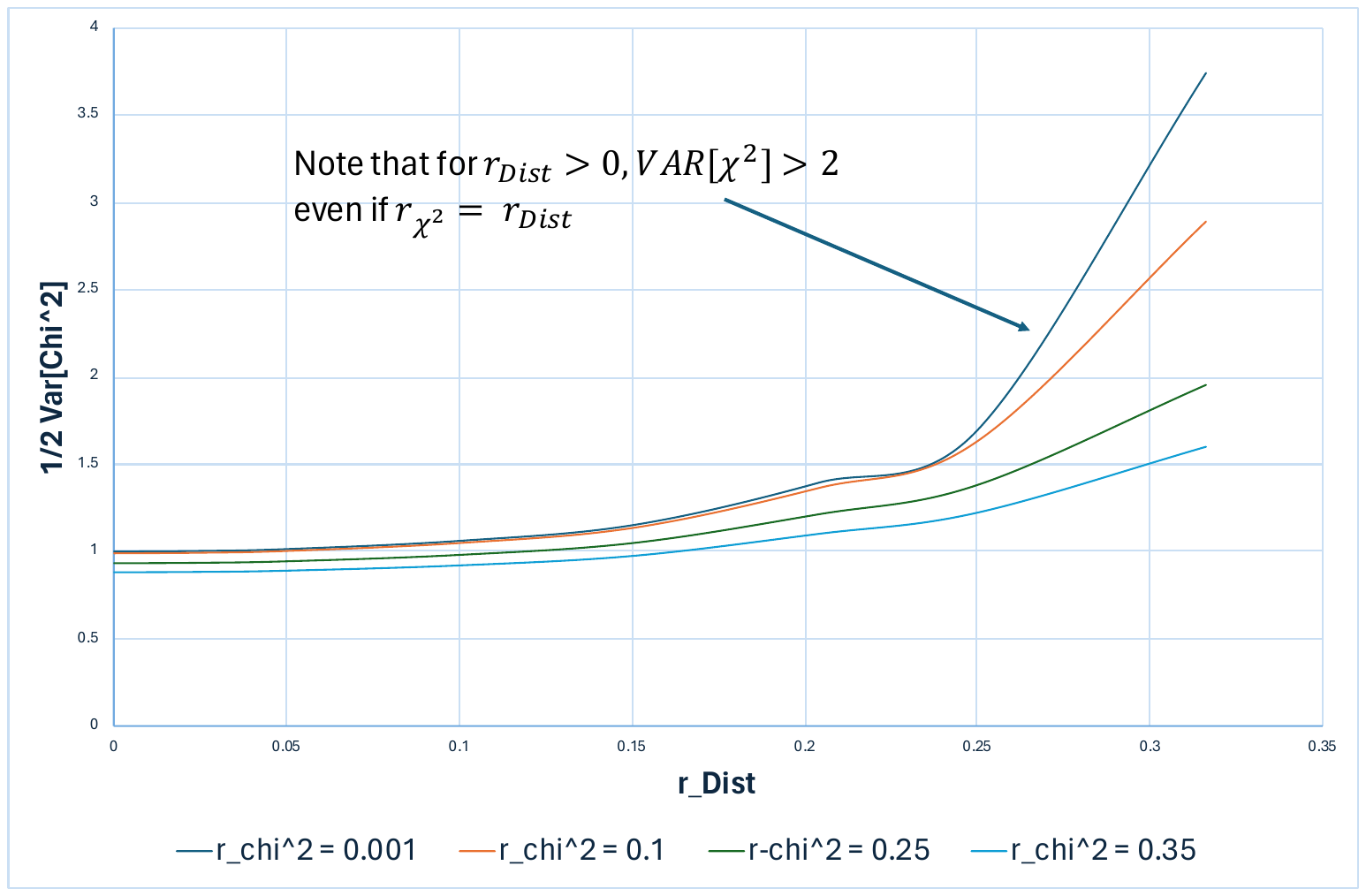}
        \caption{Graph of $Var[\chi^2]/2$ as a Function of $r_{Dist}$ for 4 different $r_{\chi^2}$ ($\sigma_{i}=\sigma_{u_i} = 1 $)}
        \label{Figure 2}
    \end{minipage}
\end{figure}

Using numerical integration we can investigate the expectation of $\chi^2$, $E[\chi^2]$, and the Variance of $\chi^2$, $Var[\chi^2]$. In Figure \ref{Figure 1} we plot the $E[\chi^2]$ as a function of $r_{Dist}$, where the $E[\chi^2]$ has been calculated at various different $r_\chi^2$. As can be seen, the expectation is a growing function of $r_{Dist}$, even if $r_{\chi^2} = r_{Dist}$. In Figure \ref{Figure 2} we plot $Var[\chi^2]/2$ as a function of $r_{Dist}$. This plot shows that the $Var[\chi^2]/2$ is a similarly increasing function of $r_{Dist}$, even as $r_{\chi^2}$ is increased.

In the case of only normally distributed statistical and t-distributed uncorrelated systematic errors, the standard deviation of the simple mean, $y_{mean} = \sum_{i=1}^N y_i/N$, is given by:
\begin{equation}
    \sigma_{Mean} \approx \frac{\sqrt{\sum_{i=1}^N \sigma_i^2+\sigma_{u_i}^2 \nu/(\nu-2)}}{N} = \frac{\sqrt{\sum_{i=1}^NE[\chi_i^2(r_{\chi^2} \rightarrow 0)](\sigma_i^2 + \sigma_{u_i}^2)}}{N}
    \label{sigma_mean}
\end{equation}
In Table \ref{Table 1}, we show how $\sigma_{Mean}$ behaves as a function of $r$ and $N$ (with $\sigma_i=\sigma_{u_i}=1$). As can be seen for all the cases of N, the ratio of $\sigma_{r_{Dist}=0.408}/\sigma_{r_{Dist}=0.0001}$ $ \approx 1.4$.

\begin{table}[!htb]
    \centering
    \begin{minipage}[t]{0.45\linewidth}
    \centering
    \begin{tabular}{|c|c|c|c|c|c|c|c|}
    \hline
\small 
$r_{Dist}$	&N=2&		N=10&		N=500 & Ratio\\
\hline
0.001& 0.995 &	 	 	 0.449& 	  	 0.064 & 1.000\\
 0.100&0.991 	 &		 0.452 &	 	 0.064 & 1.005 \\
 0.250&1.092  	 	& 0.481 &	 	 0.069&1.077 \\ 
 0.300&1.122   	& 0.504  	& 0.071 & 1.108\\
 0.408& 1.417   &	 0.637 	& 	 0.089&  1.393\\
\hline
    \end{tabular}    
    \caption{Table showing $\sigma_{mean}$ as a function of $r$ and $N$ (with $\sigma_i=\sigma_{u_i}=1$). The last column given the Ratio $\frac{\sigma_{r_{Dist}}}{\sigma_{(r_{Dist}=0.001)}}$ for the case of N=500.}
    \label{Table 1}
\end{minipage}\hfill
\begin{minipage}[t]{0.45\linewidth}
\centering
    \begin{tabular}{|c|c|c|c|c|c|c|}
    \hline
$r_{Dist}$	&N=2&			N=10&		N=500 & Ratio\\
\hline
0.001& 0.995 &		  	 0.449& 	  	 0.064 & 1.000\\
 0.100&0.991 	& 	 	 0.452 &		 0.064 & 1.004 \\
 0.250&1.092 &	   0.479 &	 	 0.068&1.069 \\ 
 0.300&1.122 &	 	 0.493 &	  0.069 & 1.087\\
 0.408& 1.417 &	 	 	 0.547 	& 	 0.076&  1.393\\
\hline
    \end{tabular} 
    \caption{Table showing $\sigma_{FIT}$ as a function of $r$ and $N$ (with $\sigma_i=\sigma_{u_i}=1$). The last column given the Ratio $\frac{\sigma_{FIT}}{\sigma_{(r_{Dist}=r_{\chi^2}=0.001)}}$ for the case of N=500.}
       \label{Table 2}
\end{minipage}       
\end{table}

The next question we want to ask is what happens if we minimize the $\chi^2$, calculated with $r_{\chi^2} = r_{Dist}$, with respect to our mean? That is, what is the standard deviation of the fitted mean, $\sigma_{FIT}$, if $r_{Dist}= r_{\chi^2}$? We show the results of this in Table \ref{Table 2} again with $\sigma_i=\sigma_{u_i}=1$.
As can be seen from the table, we have very similar behaviour to that in Table \ref{Table 1}, apart from the fact that as N increases the standard deviation  of the fitted mean, $\sigma_{FIT}$, initially starts to increase more slowly as a function of $r$, compared to the standard deviation of the simple mean.
\vspace{-0.45cm}
\section{Expectation and Variance of $\chi^2$ as a Function of $r$ for the Case of Statistical and Correlated Systematic Errors}
\vspace{-0.3cm}
Let's now consider the case of only statistical and correlated systematic errors. This is, let's consider the case of N observables each with a Gaussian statistical and M t-distributed correlated systematic errors :
$y_i =  d_i + \sigma_i z_i +  \sum_{j=1}^{M} \beta_{ij} t'_j$, where $z_i \sim N(0,1)$, and $ t'_j \sim t(0,\nu = 1/2r^2_{Dist})
$.

In the case where $r=r_{Dist} = r_{\chi^2}$, and $\sigma_i = \beta_{ij} =1 $, we obtain the data shown in Table \ref{Corr_syst}. This figure shows how the expectation of the $\chi^2$, $E[\chi^2]$, the standard deviation of the $\chi^2$, $\sigma_{\chi^2}$, the standard deviation of the simple mean, $\sigma_{\varphi_{MEAN}}$, and the standard deviation of the fitted mean, $\sigma_{\varphi_{FIT}}$, behave as function of $r=r_{Dist} = r_{\chi^2}$, the number of observables, N, and the number of correlated systematic errors, M. The behaviour is very similar to what we saw in Section \ref{Section 2} for the case of uncorrelated errors, in that the expectation and variance increase as a function of $r$, the standard deviation of the simple mean grows more quickly than the standard deviation of the fitted mean as r increases. The ratio in the last column of Figure \ref{Corr_syst}, increases to about $1.5$ in all cases compared to $1.4$ in the uncorrelated systematic error case.

\begin{figure}
    \centering
\includegraphics[scale=.57]{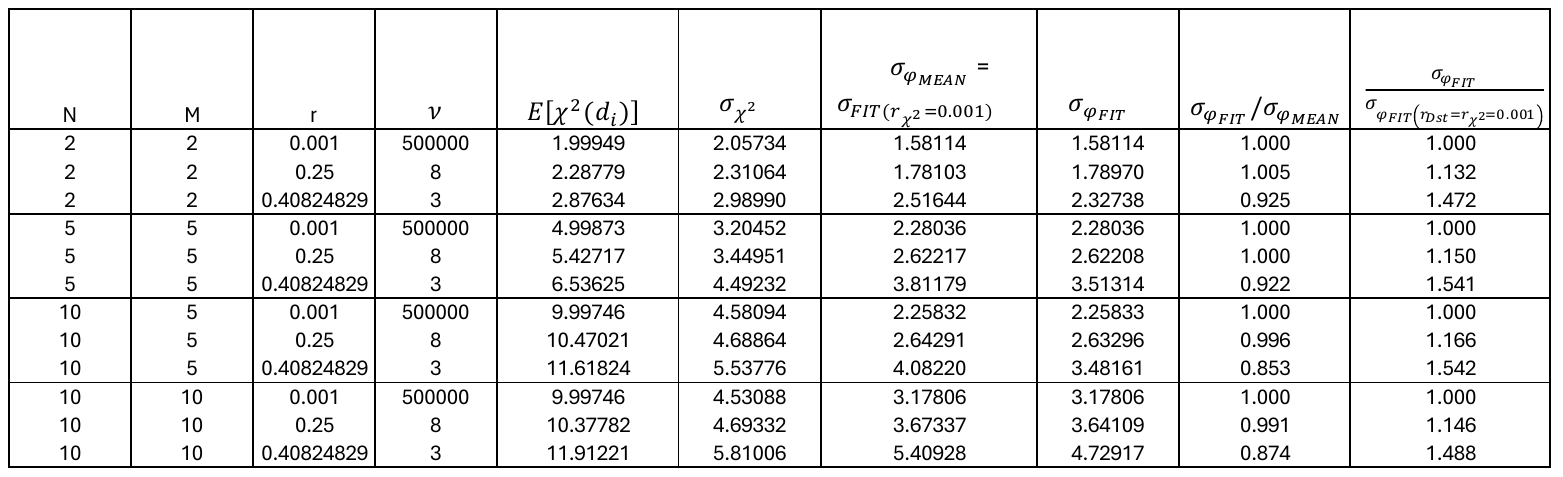} 
\end{figure}
\vspace{-6cm}
    \begin{table}
    \caption{Table showing how the expectation of the $\chi^2$, the standard deviation of the $\chi^2$, the standard deviation of the simple mean, the standard deviation of the fitted mean behave as function of $r=r_{Dist} = r_{\chi^2}$, the number of observables, N and the number of correlated systematic errors, M.} 
    \end{table}
\label{Corr_syst}
\vspace{6cm}

\vspace{-0.4cm}
\section{ATLAS W,Z Data analysis \cite{ATLAS:2016nqi}}
\vspace{-0.3cm}
This very precise data gives a strong constraint on the strange quark. However, the fit quality using the MSHT20 (NNLO) PDF set is relatively poor giving a $\chi^2 \sim 120$ . This data set consists of $61$ data points, each with $1$ statistical error, 1 uncorrelated systematic error, and 131 correlated systematic errors.
In Figure \ref{expect_W,Z} we show how the expectation of the $\chi^2$, calculated in the Gaussian limit (i.e. a $r\rightarrow 0$), varies as a function of the underlying distributional $r$, i.e $r_{Dist}$. As can be seen, the $\chi^2$ starts at 61, as expected, and increases as we increase the underlying distributional r in the simulation. It can also be seen from this Figure that $E[\chi^2]$ reaches 120 point when the underlying distribution has a $r \approx 0.4$.

Using the experimental data, we obtain the graph shown in Figure \ref{W,Z Data} when we calculate the $\chi^2$ using equation \ref{chi2_1 ru and rj t-dist}, where we have optimized with respect to $r_{u_i}$ and $r'_j$, as a function of $r_{\chi^2}$.  Once we include for this decrease in $\chi^2$ with increasing $r$ shown in Figure \ref{W,Z Data}, we can infer that some of the inflated $\chi^2$ for this data set is due to the error on errors.

\begin{figure}[!htb]
    \centering
    \begin{minipage}{.45\linewidth}
         \centering
        \includegraphics[width=1.\linewidth, height=0.25\textheight]{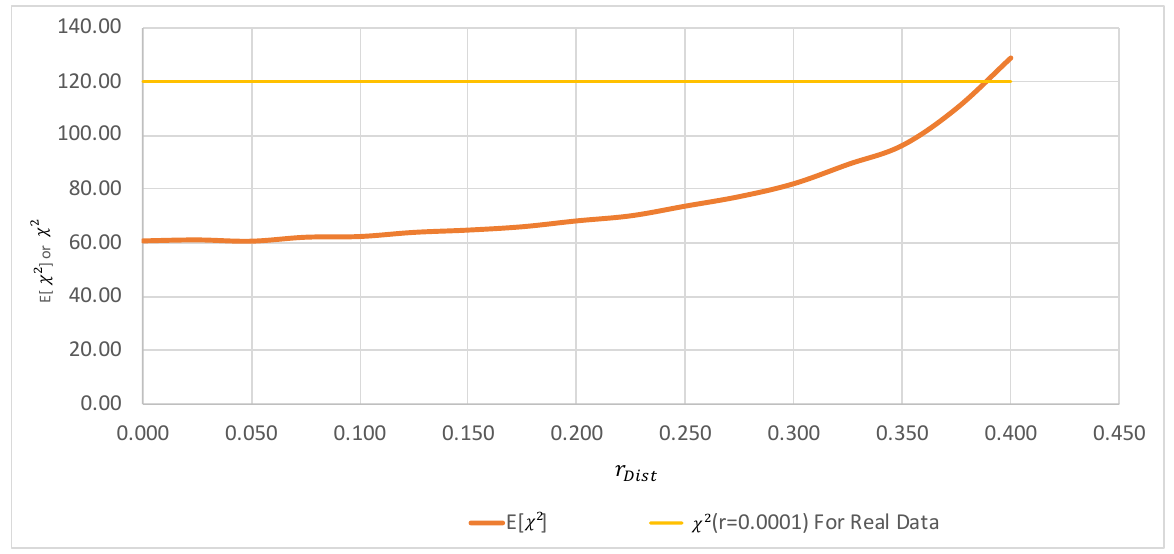}
        \caption{Graph shows the Expectation of $E[\chi^2]$, calculated using $r=0.001$, as a function of relative error, $r_{Dist}$ of the simulated underlying systematic errors.}
        \label{expect_W,Z}
    \end{minipage}\hfill
    \begin{minipage}{0.45\textwidth}
           \centering
        \includegraphics[width=1.\linewidth, height=0.25\textheight]{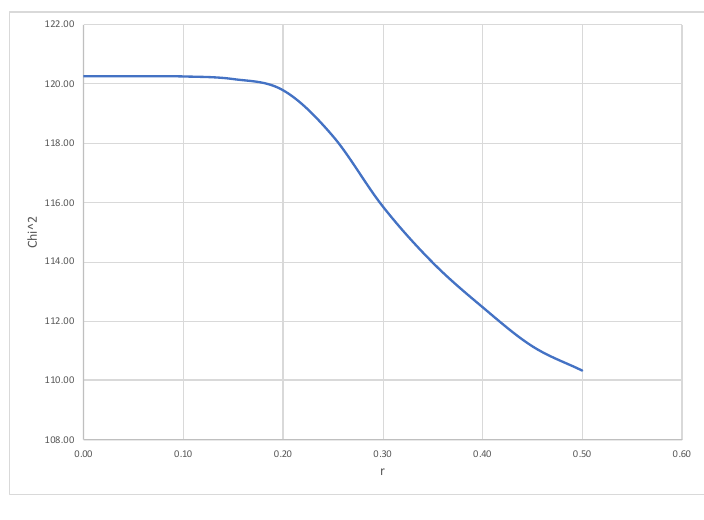}
        \caption{Graph shows the $\chi^2$ as a function of relative error, r.}
        \label{W,Z Data} 
    \end{minipage}
\end{figure}

\begin{figure}[!htb]
    \centering
    \begin{minipage}{.45\linewidth}
        \centering
        \includegraphics[width=1.\linewidth, height=0.25\textheight]{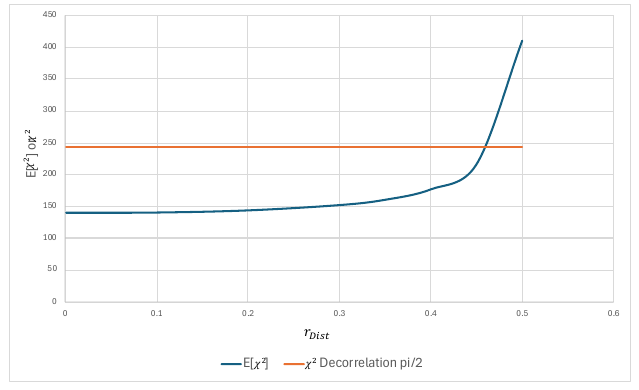}
        \caption{Graph shows $E[\chi^2]$, calculated with $r_{\chi^2}=0.00001$, where systematic errors are sampled from t-distribution with d.o.f $1/2r^2_{Dist}$. 
Line at $243.43$ is $\chi^2$ calculated using $r_{\chi^2}=0.0001$ for de-correlated data. }
        \label{expect 7Tev2}
    \end{minipage}\hfill
    \begin{minipage}{0.45\textwidth}
       \centering
        \includegraphics[width=1.\linewidth, height=0.25\textheight]{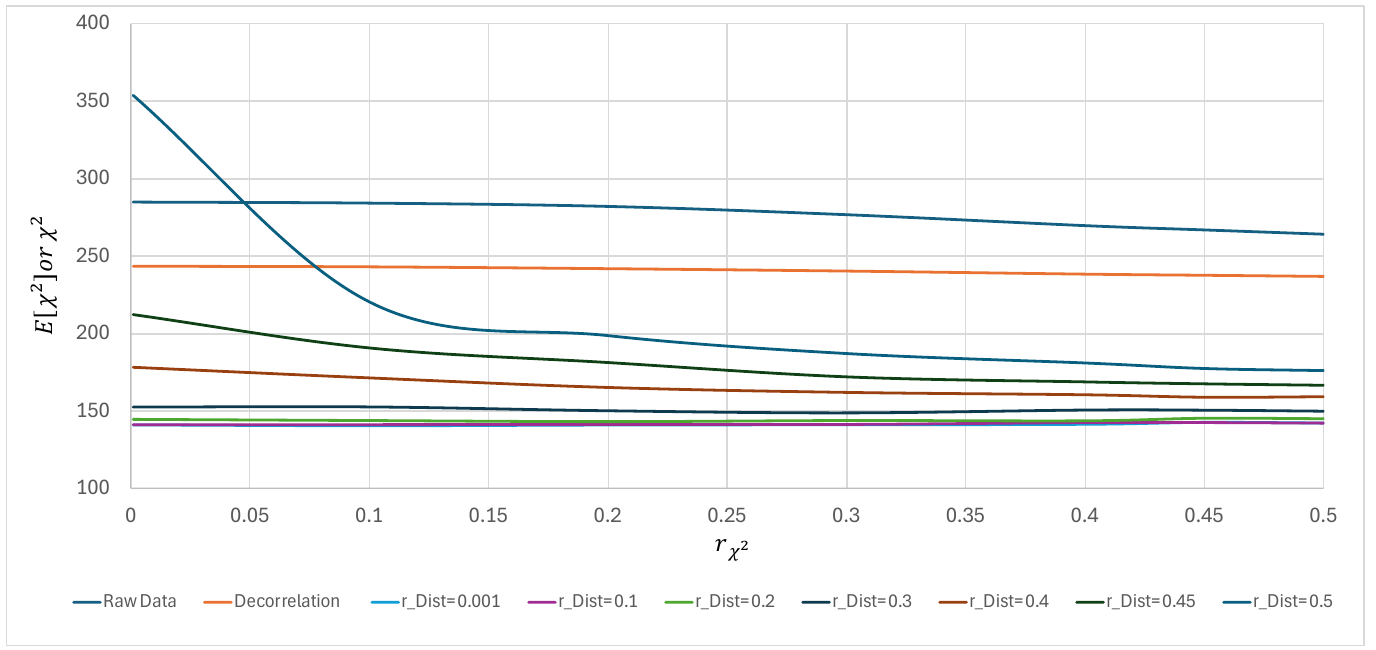}
        \caption{Graph shows the $\chi^2$ or $E[\chi^2]$ as a function of relative error, $r_{\chi^2}$. Raw data refers to just the raw data provided by ATLAS. De-correlation refers to $\chi^2$ calculated with 3 of the "two point" systematic uncertainties de-correlated. Other lines show the $E[\chi^2]$ using pseudo data produced using various $r_{Dist}$.} 
\label{expect 7TeV}
    \end{minipage}
\end{figure}
\vspace{-0.5cm}
\section{ATLAS 7 TeV Inclusive Jet Distributions Analysis \cite{ATLAS:2014riz}}
\vspace{-0.3cm}
This ATLAS data, combined with the availability of NNLO corrections provides constraints on the Gluon PDF at high x. The data set consists of 140 data points, each with 1 correlated systematic error and 70 correlated systematic errors. Using the MSHT20 NNLO PDF set, the fit quality to this data set is relatively poor with a $\chi^2 \approx 280$. In order to improve the fit quality MSHT20 \cite{Bailey:2020ooq} use a de-correlation process which reduces the $\chi^2 \approx 243$.

In Figure \ref{expect 7Tev2}, we show a similar graph to that in Figure \ref{expect_W,Z} for this data set. The graph shows the expectation of $\chi^2$, $E[\chi^2]$, calculated with $r_{\chi^2}=0.00001$, where the systematic errors are sampled from t-distribution with d.o.f $1/2r^2_{Dist}$. As can be seen $E[\chi^2]$ crosses the $\chi^2=243$ line at an $r_{Dist}$ of about $0.45$. 
In Figure \ref{expect 7TeV}, we show how the $\chi^2$ varies as a function of $r_{\chi^2}$ for the cases of just the raw data, and for the case of the MSHT20 de-correlation procedure (labelled "Decorrelation").
Also in Figure \ref{expect 7TeV}, we show how the expectation of $\chi^2$, $E[\chi^2]$, behaves as a function of $r_{\chi^2}$ for different values of $r_{Dist}$.  Suggestive as it is, making the assertion that this data has an $r \approx 0.45$, would neglect not only the decreasing behaviour of $\chi^2$ as a function of $r$, but also the theoretical uncertainties and the choice of de-correlation process used.

\vspace{-0.3cm}
\section{Conclusions}
\vspace{-0.3cm}
In this document we have shown how we can incorporate Errors on Errors into the calculation of a $\chi^2$.
We have also shown that the Expected $\chi^2$ and Variance of $\chi^2$ increase as the relative errors of the systematic errors increase. We have noted that for both data sets analysed $r \approx 0.4$. We have also observed that the ratio of the expected standard deviation of the mean, using $r_{\chi^2} = 0.001$ and $r_{Dist}=0.4$, to that calculated using $r_{\chi^2} = 0.001$ and $r_{Dist}=0.001$ is approximately $1.2-1.5$. This is suggestive of using a Tolerance, $T^2$, in the region of  $1.5-2$ in these test cases.

\end{document}